# Scenarios and branch points to future machine intelligence

Koichi Takahashi[*12]
ktakahashi@riken.jp

[*1] RIKEN Center for Biosystems Dynamics Research

[*2] Keio University Graduate School of Media and Governance

We discuss scenarios and branch points to four major possible consequences regarding future machine intelligence; 1) the singleton scenario where a single super-intelligence acquires a decisive strategic advantage, 2) the multipolar scenario where the singleton scenario is not technically denied but political or other factors in human society or multi-agent interactions between the intelligent agents prevent a single agent from gaining a decisive strategic advantage, 3) the ecosystem scenario where the singleton scenario is denied and many autonomous intelligent agents operate in such a way that they are interdependent and virtually unstoppable, and 4) the upper-bound scenario where cognitive capabilities that can be achieved by human-designed intelligent agents or their descendants are inherently limited to the sub-human level. We identify six major constraints that can form branch points to these scenarios; (1) constraints on autonomy, (2) constraints on the ability to improve self-structure, (3) constraints related to thermodynamic efficiency, (4) constraints on updating physical infrastructure, (5) constraints on relative advantage, and (6) constraints on locality.

**Keywords:**
AI alignment, thermodynamics of computation, intelligence explosion, machine intelligence, technological singularity

## 0. Preface

This article is an English translation of a paper originally presented in Japanese at the 32nd annual conference of Japanese Society for Artificial Intelligence in June 2018[Takahashi 18a], which was later revised and published on the Journal of Japanese Society for Artificial Intelligence in November 2018[Takahashi 18b].

## 1. Introduction

In this paper, we attempt to classify the possible outcomes of the development of machine intelligence into several scenarios extending into the relatively distant future, and to identify the major branch points assumed to exist along the way to each of these scenarios.

This paper makes the following assumptions to focus on the main issues of identifying major branch points. First, any technology not prohibited by the laws of physics will be realized, provided that sufficient resources are invested. Second, among the technologies not physically prohibited, those expected to have economic or other benefits as a result of an economically rational investment of resources will be realized by a certain entity. Here, note that there are dependencies between technologies (e.g., if the microphone had not been invented, the telephone would not have been developed). Third, technological diffusion occurs at a constant rate so that the invented technology is shared after a certain time. Fourth, temporary power imbalances between competing entities converge to an equilibrium after a certain amount of time. However, stochastic fluctuations can trigger irreversible bifurcations between scenarios, which pose uncertainty factors as the accidental outcomes can be historically fixed.

## 2. Scenarios

This section introduces the scenarios that can be envisioned through the long-term development of machine intelligence technology.

### 2.1 The Singleton Scenario

This is a scenario in which the rate of self-improvement of an intelligent agent that recursively updates itself as if it has no upper limit until it gains a decisive strategic advantage [Bostrom14].

Bostrom refers to decisive strategic advantage as 'the level of technological and other advantages sufficient to enable complete world domination'[Bostrom 14]. In this paper, we give it a more specific definition as having effective countermeasures against all possible moves by the opponents. Hegemony obtained is difficult to overturn in resource acquisition and in competition with other agents.

### 2.2 The Multipolar scenario

In this scenario, the performance of agents stagnates before they achieve a decisive strategic advantage due to external factors such as the establishment of international treaties based on the recognition of the danger of self-improving machine intelligence and/or the danger of hegemony by a singleton. However, possible occurrence of a singleton due to changes in power relations among the actors involved in the development of machine intelligence or the machine intelligence themselves, and other uncertainty factors such as terrorism is not ruled out [Bostrom14].

## 2.3 The Ecosystem scenario

This scenario assumes that there is a limit to the performance improvement of machine intelligence before they can reach a decisive strategic advantage, resulting in a network of coexistent and interdependent agents. Such an "AI ecosystem" cannot be shutdown, as is the case with the Internet and power grids today because human activities heavily depend on them, and their overall behavior will be unpredictable if the behavior of individual agents or interdependence network of the agents is complex to a certain degree [Yamakawa17].

## 2.4 The Upper-bound scenario

This is a scenario in which there is a fundamental upper limit to the development of their capabilities of human-engineered machine intelligence, and that they will not acquire the ability to operate autonomously without instructions from humans.

## 3. Constraints and Branch points

In this section, we list the constraints that would determine which of the machine intelligence scenarios will realize.

## 3.1 Constraints related to internal structure

### (1) Constraints on autonomy

Autonomy is closely related to task versatility, the ability to cope with a wide range of situations, including exceptional circumstances and environmental changes. Many of the arguments that fall into the upper bound scenario claim that there are insurmountable obstacles that prevent the cognitive capabilities of an engineered cognitive architecture from reaching human-like levels of task versatility. However, this is not physically forbidden if we take the materialistic standpoint that systems that are computationally identical to the neural connections in the human brain would have identical cognitive capabilities. However, it has not yet been verified what kind of further technological developments are required to design an internal structure with sufficient performance and how much research and development costs should be spent for it. In addition, discussions on the contribution to task versatility of the emergent dynamics generated by interaction with the environment and the body, and of the cognitive functions acquired through learning, are far from conclusive.

### (2) Constraints on the ability to improve self-structure

Whether an agent can acquire capability to improve their own internal structure is an important issue. That is, whether an agent can generate new structural information of an agent with a higher response speed or cognitive ability than its own assuming that the available computational power is constant, starting from the structural information of its own or that of an external agent with the same ability. This might seem a far-reaching discussion, but the fact is that it is essentially an engineering problem involving the modularity of the internal structure. If the architecture has modularity and hierarchy (*i.e.*, near decomposability [Simon96]), the search problem of improving the entire architecture of improved performance or capacity can be divided into a set of sub-problem of orders of magnitude smaller search space. Such architecture search sub-problems will include improvement of internal structure of a module that constitutes the entire architecture and the entire architecture itself which designs how these sub-modules are connected to each other. Nonetheless, the hypothesis that it is possible to design intelligent agents with human-level or higher capabilities as a near decomposable system has not yet been tested (not to mention, near-decomposability of human brain connectome). The upper limit of the improvement speed will be highly related to the execution speed of the inference cycle that predicts the performance improvement by the design change. An evolutionary computational approach using genetic algorithms would also be possible, but again, the time required for simulation defines the upper bound on the improvement speed.

## 3.2 Constraints related to physical properties of computing elements

### (3) Constraints related to thermodynamic efficiency

There is a thermodynamic upper limit to the possible amount of computation that can be drawn out from a given amount of energy. The second law of thermodynamics dictates that logically irreversible operations such as erasure of information involve an increase in thermodynamic entropy equivalent to $k_B$ T ln 2 or more per bit (the Landauer limit, where $k_B$ is Boltzmann limit and T is absolute temperature). This is approximately $2.87 \times 10^{-21}$J at room temperature (300K). The switching energy of a modern computer is about $10^{-17}$J [Theis17], about 10,000 times as close as this limit. Information loss per a floating-point arithmetic operation is about N bits (ideal reversible gate-based implementation) to N log N bits (more realistic estimate) for a precision of N. If N = 10 bits is assumed, the loss is roughly about $10^{-18}$J. On the other hand, the human brain conducts about $10^{16}$ calculations per second with an energy of about 10W. Comparing these numbers, it could be said that physical laws do not prohibit development of a commodity device that conducts an equivalent amount of computation. At the same time, this consideration also tells that a large amount of energy will be required to achieve performance far beyond those of humans on conventional digital computers.

This limit does not apply locally to the kind of computations that do not lose information, such as quantum computation in which the quantum state is not destroyed or to molecular machines including DNA computing which utilizes reversible processes. However, when and how such types of computing devices would be available is uncertain.

### (4) Constraints on updating the physical infrastructure

Every physical phenomenon has a time constant associated with it. In order for an agent to significantly improve its capabilities and response time without changing the amount of resources it uses, it needs to update its physical infrastructure, except in cases where the architecture is only insufficiently optimized. The use of new physical phenomena or combinations of them requires the generation of hypotheses about the performance improvement and their verification by physical experiments, and the time required for the verification depends on the physical time constant of the subject matter. When using simulation for knowledge discovery, it may be faster than the physical time constant if it is related to emergent properties that do not involve new physical phenomena. However, searching for new knowledge that involves inherent properties of physical elements, simulations generally take longer than real time because it requires a lower level and finer granularity of computation than the physical phenomena of interest belong to (realist approach). When an unknown process is assumed and explored through model validation at the same level, it is necessary to test many possible hypotheses by physical or virtual experiments, which takes even longer time. These are factors that limit the speed of self-improvement based on the update of the computing elements.

## 3.3 Multi-agent constraints

### (5) Constraints on relative advantage

The constraint on the time constant provides an additional precondition for relative advantage among agents. If there is no significant difference in the order of magnitude in the physical capacities such as the mass and amount of available energy that each agent has under its control, capabilities to make predictions on other agent's behaviors and their consequences have a greater emphasis. In a multi-agent situation where agents are interacting and only incomplete information is available about the environment and other agents, an effective way to gain an advantage over other agents is to gain an ability to predict the actions of other agents and their consequences more quickly than others. To achieve a dominant position over other agents using such predictive capability, it is necessary to maintain the ability to make predictions on other agents' behaviors and their consequences. This prediction needs to be for a longer time scale than that of both the response time of the target agent to external perturbations and the physical and network communication latency of the predicting agent and the target agent.

This consideration leads to two important consequences. First, in a multi-agent situation in the physical world, a slight advantage in prediction ability and response time over other agents is not immediately a sufficient condition for behavioral advantage. From the viewpoint of computational complexity, many problems require polynomial or exponential computation, rather than linear, with respect to the scale of the problem such as the number of variables. So, generally only logarithmic utility can be obtained with respect to the increase in computational power. Therefore, superiority in predictive power does not immediately mean being able to reject attacks from other agents, but overwhelming superiority at exponential scales will be necessary to surely gain relative advantage. Second, on the other hand, if a resource advantage gained by chance under stochastic conditions happens to lead to a temporary advantage close to decisive and if the agent in question makes technological or resource-acquisition progress that transforms the relative advantage into a decisive strategic advantage during the relaxation time it takes for the entire system including the other agents to catch up and return to the equilibrium, then it may result in a scenario bifurcation (so-called 'frozen accident' discussed in e.g. [de Duve05]. Therefore, scenarios in which a decisive strategic advantage is gained due to stochastic fluctuations is possible.

### (6) Constraints on locality

The boundary between the agent's self and others in physical space is determined by the arrangement of the sensor and actuator systems under the agent's control. If we put the speed of light as a constant, the spatial distance between the main computational elements of the agent defines the upper limit of the agent's response speed. (This is similar to the way that the cognitive response time of the brain of about 100 milli seconds has its basis on the physical size of the human brain and the speed of neural transmission.) According to the information integration theory (IIT), in order for conscious experience to be experienced as a unified whole, rather than as a collection of separate parts, there must be an informational coupling between elements measured by the amount of mutual information [Tononi16]. When an agent has a spatial extent, the speed at which it makes decisions and responds based on the experience obtained through such informational integration is limited by the delay caused by the speed of light associated with the communication between elements. When the communication with the sensor and actuator systems is unidirectional, the spatial distance does not directly limit the response speed inside the agent, but it does limit the response speed in the form of delays in acquiring information about the environment and other agents and in action of the agent. These limits, combined with the constraints on relative advantage due to limits on response speed described in the previous section, set demands on the localization of the spatial extent of the agent's capabilities and, indirectly, on the amount of distributed computational resources available.

An agent can attempt to avoid the constraints on locality and reduce the response time to local events at a remote place by adopting an asynchronous consensus process between an arbitrary number of copies of itself or other types of agents under its influence deployed in advance and communicating with them based on predictions of future changes in the situation. However, in order for multiple agents to share a decision-making process based on unified empirical information, communication generally involves replication of information. Brewer's CAP theorem for distributed systems shows that there is a trade-off among consistency, availability, and partition-tolerance for information replication between nodes, and that in general only up to two of these can be guaranteed simultaneously [Lynch02]. It is also known that, in a distributed consensus process, a single process failure is sufficient to make consensus making in finite time impossible (FLP impossibility) [Fischer85]. In practice, the difficulties indicated in these theorems are not unavoidable in many cases if one can wait for a sufficiently long time for failure recovery. Nevertheless, what these theorems indicate here is that there is an upper limit to the response speed that a distributed system can guarantee. Considering these considerations together with the speed of light, it suggests that there are upper limits on the computational power within a certain response time that can be utilized to pursue relative advantage not only in the case of a single agent but also in the case of multiple agents communicating with each other. Therefore, using the additional computational resources obtained by generating multiple instances cannot give the agent the ability to pursue relative advantage indefinitely.

## 4. Scenarios and branch points

Using the constraints discussed in the previous chapter, this section examines the branch points to the scenarios classified in Chapter 2. As mentioned in the introduction, the purpose of this paper is not to predict the future, but to identify the branching points for the scenarios.

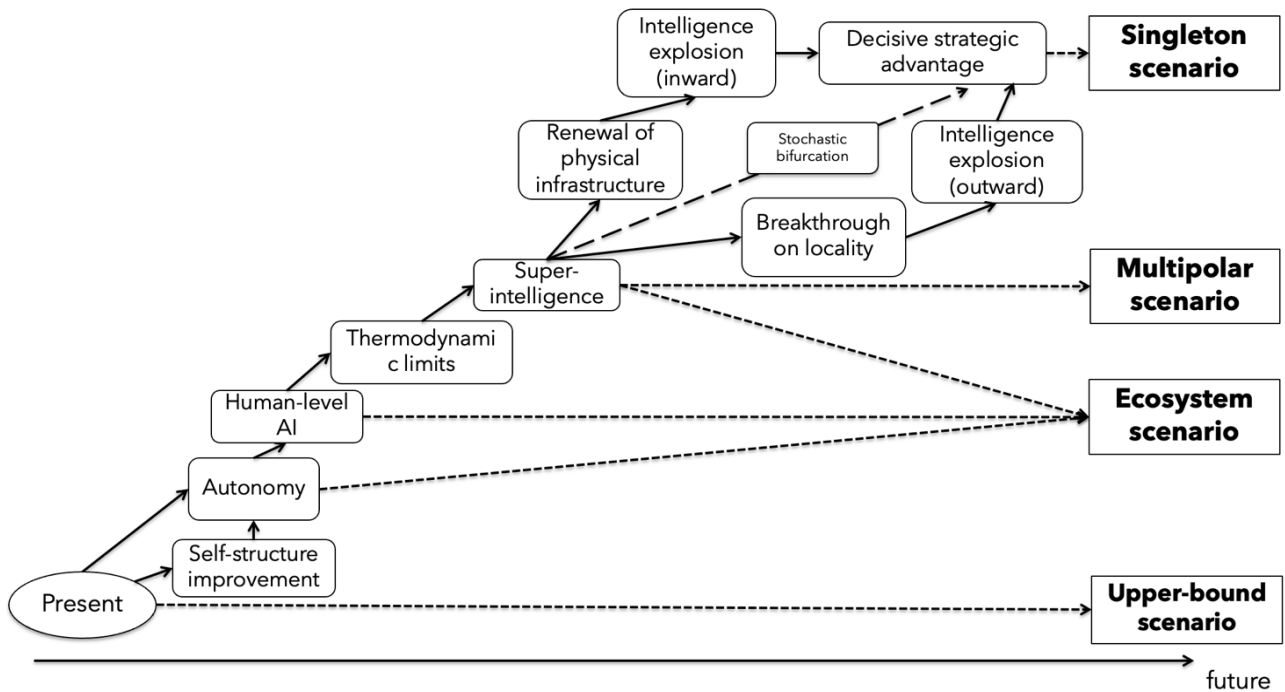

Fig. 1 Scenarios and branch points

While *the constraints on autonomy* do not prohibit machine intelligence from acquiring cognitive abilities equivalent to or greater than those of humans in the future, whether such technology can be developed at a realistic cost is still unclear. The other possibility is that a manually designed machine cannot achieve human-like cognitive performance, but it is possible to design a machine with *the ability to improve self-structure,* resulting in beyond-human intelligence. These two paths lead to a bifurcation between *the upper-bound scenario* and the other scenarios.

*The singleton scenario* is a scenario in which all the limitations described in the previous section are breached or avoided and a single agent secures a decisive strategic advantage. A crucial question here is what determines the level of cognitive ability required to secure the decisive strategic advantage. It is not enough to simply outperform all other agents in terms of reasoning ability and response time, but it is also necessary to have the ability to predict the entire open system, including disturbances caused by environmental factors. In absolute terms, the first hurdle is the prediction of the environment. In relative terms, it is necessary to have orders of magnitude more computational power than any other agent in order not to be outsmarted by other agents and have its advantage threatened. In order to reach such a situation, the agent must either consolidate its advantage, such as controlling the majority of computational resources (*i.e. outward* intelligence explosion), or secure the ability to expand its initial advantage exponentially by updating its own *physical infrastructure* (*i.e. inward* intelligence explosion). In the outward intelligence explosion scenario, both the *constraints on relative advantage* and the *constraints on locality* need to be breached, which requires many stringent conditions to be met. In the inward intelligence explosion scenario, the critical issue is under what conditions the machine intelligence that updates its own *physical infrastructure* will occur.

Generally, improved prediction of state changes in the environment is advantageous because it provides reduced uncertainty, which is useful for raising the probability of achieving one's goals. Therefore, autonomous agents are generally considered to have a propensity to acquire resources that improve their ability to predict. Such resources will include computational resources and access to sensor/action systems. Both the *singleton scenario* and the *multipolar scenario* do not rule out the possibility of occurrence of a singleton. The branch between these two scenarios may rather depend on whether the propensity to acquire resources can be artificially limited by design.

The bifurcation between *the ecosystem scenario* and *the upper-bound scenario* will occur depending on whether and how well the machine intelligence would acquire the ability to autonomously execute the cycle of perception, judgment, and action decision without human instruction (*constraints on autonomy*). Once a machine intelligence with an advanced level of autonomy is realized, this technology will soon spread to virtually any domain in society to automate various tasks due to a vast economic rationale for using it. If the utility of automated machines is measured by their autonomy, that is, the amount of time they can act without human directions, then autonomy will be pursued as long as the technical and economic costs are reasonable. In the present day, many human activities are already based on computer networks, and autonomous machine-intelligent agents will not only become interdependent with human society in a relatively short period of time, but will also establish interdependent and mutually complementary relationships among agents.

*The ecosystem scenario* actually gives rise to several sub-scenarios depending on the level of intelligence that is composed of (*i.e.,* sub-human-level, human-level, or super-human-level). In this sense, it would be possible to view *the multipolar scenario* as a subcategory of the ecosystem scenario which is composed of superintelligences. The only difference between the superintelligence ecosystem scenario and the multipolar scenario is the potential acquisition of a decisive strategic advantage.

Fig. 1 shows the relationships between scenarios and branch points in the form of a directed graph starting from the present to each of the consequential scenarios.

## 5. Conclusion

In this paper, we discussed the possibility that the long-term development of machine intelligence may depend on various technological, economical, political, and physical constraints that can set the upper limits in their capability levels. Such constraints may give rise to the scenario branches in the following order: *the upper-bound scenario*, *ecosystem scenario*, *multipole scenario*, and *singleton scenario*. In addition to architectural-level issues such as the realization of a cognitive architecture with a high degree of *autonomy* and the ability to improve its own structure, we also discussed the difficulty of establishing a *relative advantage* in multi-agent situations, in addition to some physical constraints such as the *thermodynamic efficiency* of computation and the upper limit set by the speed of light.

Although technological singularity is not a main subject here, we would like to spend some words on it to attempt to position it in the context of scenarios we have structured in this paper. Technological singularity is closely related to the notion of inward intelligence explosion discussed in the previous chapter. Vernor Vinge described four possible scenarios: 1) the emergence of superintelligence through the "awakening" of computers, 2) the emergence of superintelligence through the "awakening" of computer networks, and 3) the fusion of machine and human intelligence through technologies such as BMI (Brain-Machine Interface), and 4) the development of biotechnology, resulting in human super-intelligence [Vinge93]. Our discussion in this paper mostly concerns Vinge's scenario 1. Vinge's scenarios 3 and 4 start with, unlike in his scenario 1, human-level intelligence, so it will remain within the realm of the ecosystem scenario unless it proceeds to *self-renewal of physical elements* to break through the *thermodynamic limit* at some point. Scenario 2 that concerns the awakening of the computer network would be strongly constrained by the constraints on localization, and unless this is circumvented in some way, it will remain either an *upper-bound scenario* or an *ecosystem scenario*, with severe limits imposed on its ability to predict and manipulate real-world events due to its limited response time.

Finally, it should be noted that if the constraint on *thermodynamic efficiency* and *the constraint of localization* that is closely related to the upper limit set by the speed of light are withdrawn, the only constraints remaining before the branch to the *singleton scenario* will be the *self-renewal of physical elements* and the *relative advantage*. There is a possibility that the thermodynamic efficiency may not pose a fundamental limit. For example, the Landauer limit would not apply to quantum computation and DNA computation when these are considered as reversible computations [Toffoli05] (see also 3.1(3)). On the other hand, the upper limit set by the speed of light appears to be much more stringent. So far, only physical phenomena with energies below the TeV scale have been experimentally explored, and no theory has yet been discovered that can describe all forces in a unified manner, so we do not know if the speed of light is the real limit. Nevertheless, no method has been currently shown to transmit information at speeds faster than the speed of light, as far as engineering can possibly handle.

## Ackowledgements

The author would like to thank Hirotaka Osawa, Hiroshi Yamakawa, and Makoto Taiji for their technical advice. Naoya Arakawa gave us a great help in translating the original article written in Japanese to English. We are also grateful to Hitomi Sano for her help in editing and preparing the figures. Part of this research was supported by KAKENHI Grant Number 17H06315, Grant-in-Aid for Scientific Research on Innovative Areas, Brain information dynamics underlying multi-area interconnectivity and parallel processing, the MEXT Post-K exploratory challenge 4 "Big Data Analysis of the Brain, Whole Brain Simulation and Brain-based Artificial Intelligence Architecture," and JST-RISTEX HITE "Co-creation of Future Social Systems through Dialogue between Law, Economics, Management, and AI/Robotics Technologies.

## References

[Takahashi 18a] K. Takahashi: Scenarios and branch points to future machine intelligence, The 32nd Annual Conference of the Japanese Society for Artificial Intelligence, Kagoshima, June 2018, (1F3-OS-5b-03)
https://doi.org/10.11517/pjsai.JSAI2018.0_1F3OS5b03

[Takahashi 18b] K. Takahashi: Scenarios and branch points to future machine intelligence, Journal of Japanese Society for Artificial Intelligence, Vol. 33 No. 6 p867-871
https://doi.org/10.11517/jjsai.33.6_867

[Bostrom 14] N. Bostrom: Superintelligence: Paths, Dangers, Strategies, OUP Oxford, 2014.

[Yamakawa 19] H. Yamakawa: Peacekeeping Conditions for an Artificial Intelligence Society. Big Data and Cognitive Computing, 3(2), 34, 2019.

[Simon 96] H. A. Simon: The Sciences of the Artificial, MIT Press, 1996.

[Watabe 16] M. Watabe, T. Tsuzuki, K. Kaizu, K. Takahashi: Accelerating Science by Artificial Intelligence, The 30th Annual Conference of the Japanese Society for Artificial Intelligence, Kitakyushu, June 2016, 2E5-OS-12b-2, https://doi.org/10.11517/pjsai.JSAI2016.0_2E5OS12b2

[de Duve 05] C. de Duve: Singularities: Landmarks on the Pathways of Life, Cambridge University Press, 2005.

[Theis 17] T.N. Theis, H.-S. P. Wong: Computing in Science and Engineering, pp.41-50, 2017.

[Tononi 16] T. Giulio, B. Melanie, M. Marcello, K. Christof: Integrated information theory from consciousness to its physical substrate, Nature Reviews Neuroscience, pp.450–461, 2016.

[Lynch 02] N. Lynch, S. Gilbert: conjecture and the feasibility of consistent, available, partition-tolerant web services, ACM SIGACT News, pp. 51-59, 2002.

[Fischer 85] Fischer, Michael J., Nancy A. Lynch, and Michael S. Paterson. Impossibility of distributed consensus with one faulty process. Journal of the ACM (JACM) 32.2 pp. 374-382, 1985.

[Vinge 93] V. Vinge: The Coming Technological Singularity: How to Survive in the Post-Human Era, 1993. https://edoras.sdsu.edu/~vinge/misc/singularity.html

[Toffoli 05] T. Toffoli: Reversible computing, Lecture Notes in Computer Science book series, Springer, 2005.